\documentclass[
    ,final            
  ]
  {aipproc}

\layoutstyle{6x9}
\newfont{\sfsl}{cmssqi8 scaled 1300}
\newfont{\sfsls}{cmssqi8 scaled 1150}
\newcommand{\gcs}{{\sfsl HIFLUGCS}}
\newcommand{\gcss}{{\sfsls HIFLUGCS}}

\begin{document}

\title{Cosmic Structure Traced by Precision Measurements
of the X-Ray Brightest Galaxy Clusters in the Sky}

\author{Thomas H. Reiprich}{
  address={Department of Astronomy, University of Virginia, PO Box 3818,
  Charlottesville, VA 22903-0818}
  ,altaddress={thomas@reiprich.net, http://www.reiprich.net} 
}

\author{Craig L. Sarazin}{
  address={Department of Astronomy, University of Virginia, PO Box 3818,
  Charlottesville, VA 22903-0818}
}

\author{Joshua C. Kempner}{
  address={Harvard-Smithsonian Center for Astrophysics,
  60 Garden Street, MS-67 Cambridge, MA 02138}
}

\author{Michael F. Skrutskie}{
  address={Department of Astronomy, University of Virginia, PO Box 3818,
  Charlottesville, VA 22903-0818}
}

\author{Gregory R. Sivakoff}{
  address={Department of Astronomy, University of Virginia, PO Box 3818,
  Charlottesville, VA 22903-0818}
}

\author{Hans B\"ohringer}{
  address={Max-Planck-Institut f\"ur extraterrestrische Physik, PO Box 1312,
  85741 Garching, Germany}
}

\author{J\"org Retzlaff}{
  address={Max-Planck-Institut f\"ur extraterrestrische Physik, PO Box 1312,
  85741 Garching, Germany}
}

\begin{abstract}
The current status of our efforts to trace cosmic structure with $10^6$ galaxies
(2MASS), $10^3$ galaxy clusters (NORAS\,II cluster survey), and precision
measurements for $10^2$ galaxy clusters (\gcss) is given. The latter is
illustrated in more detail with results on the gas temperature and metal abundance
structure for $10^0$ cluster (A1644) obtained with XMM-Newton.
\end{abstract}

\maketitle


\vspace{-0.1cm}
Galaxy clusters have been very important tools to study cosmic structure and
determine cosmological parameters. Even moderately sized samples yield
competitive statistical constraints, e.g., \gcs\ (Fig.~1), \cite{rb01}. However,
accuracy is currently limited by systematic uncertainties. Luckily major
improvements are now feasible, e.g., by taking advantage of new multiwavelength
data and especially Chandra and XMM-Newton X-ray observations. We've started a
project to tackle systematic uncertainties in flux, temperature, gas and
gravitational mass estimates by detailed analyses of the 63 X-ray brightest
galaxy clusters in the sky (\gcs). Many of these clusters have already been
observed by Chandra and XMM-Newton. Data are accumulating in the archives.
The first out of further nine approved Chandra observations proposed by us is
being carried out today (2002-12-16).

\begin{figure}
  \includegraphics[height=.67\textheight, angle=270.]{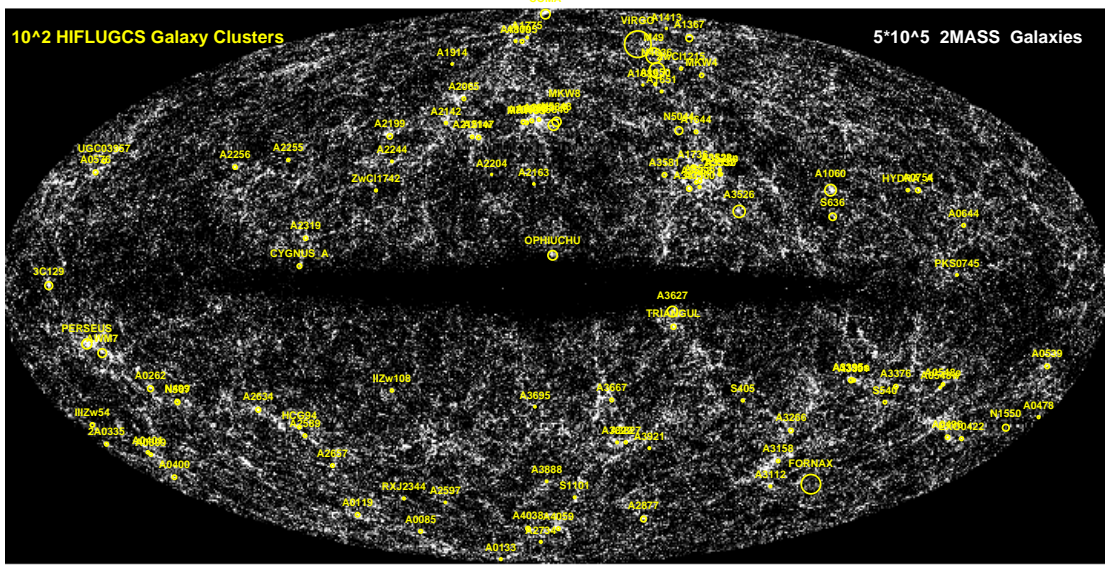}
  \caption{\gcss\ and other local X-ray galaxy clusters shown in Galactic
  coordinates on top of the distribution of half a million 2MASS galaxies.
  Qualitatively both populations seem to nicely trace an underlying matter
  distribution with clusters being located at the peaks.\vspace{-0.6cm}}
\end{figure}

The NORAS\,II cluster survey contains about 800 X-ray selected galaxy clusters
(B\"ohringer, Retzlaff et al., in prep.). In order to test for systematic
effects caused by different selection techniques --- which must be well
understood to obtain precise cosmological constraints ---  we started
correlating the
NORAS\,II clusters with the 2MASS extended source catalog (the latter shown in
Fig.~1, \cite{jcc00}) and color
selected point source catalog. Furthermore galaxy overdensities at cluster
positions will be determined in order to estimate richness--mass relations and
mass-to-light ratios.

In the following we illustrate possible improvements from the X-ray side
on an especially tough example. XMM-Newton observations of
A1644 show a very complicated surface brightness distribution on all
scales (Reiprich et al., in prep.). A main clump and a smaller sub clump are
easily identified. The emission surrounding both core regions is highly
non-spherical. And the core region of the main clump itself contains a displaced
core-within-a-core. How much do X-ray flux, gas,
and gravitational mass estimates based on precision measurements differ from
simple estimates where, e.g., the whole cluster is treated as a spherical cow
as might be done if only ROSAT All-Sky Survey data were available? How much
do mass estimates differ when only a broad beam overall gas temperature estimate
is available?

The X-ray flux ($f_X$) ratio between the main and sub clump is about 3:1. This
means instead of one cluster in a flux-limited sample with $f_X\approx 4\times
10^{-11}$\,erg/s/cm$^2$ one actually has two clusters with about $f_X\approx 3$
and $1\times 10^{-11}$\,erg/s/cm$^2$ each. This is quite an important
difference, especially for construction of luminosity and mass functions!

The intraclucter gas temperature is a fundamental observable for a
mass determination based on the hydrostatic assumption. One first step of
refinement compared to broad beam temperature estimates is the construction of
radial temperature
profiles for the main and sub clump. To our surprise we found that each of the two
temperature profiles appears very similar to temperature profiles of relaxed,
apparently undisturbed
clusters: a drop in the center to about 1/2 to 1/3 of the ambient gas
temperature, an isothermal structure in the outer parts, and weak
indications for a slight temperature drop in the very outermost regions
accessible. This appears to be good news: this
cluster may not be as complicated as it seemed, the temperature structure may
not be affected by the interaction of the two sub clumps (which are located at
about the same redshift). Note, however, that simple broad beam temperature
estimates would still be biased low compared to the ambient gas temperature
due to cool emission in the dense (high emissivity) centers. For instance, the
temperature estimate in a large annulus around the main clump --- taken
as the ambient temperature --- is a factor of 1.1--1.15 higher than
a broad beam temperature estimate including both clumps. Since $M\propto
T^{1.5-2.0}$ \cite[e.g.,][]{frb00} this factor translates into a factor
of 1.15--1.32 in a mass estimate!
 
\begin{figure}
  \includegraphics[height=.45\textheight, angle=0.]{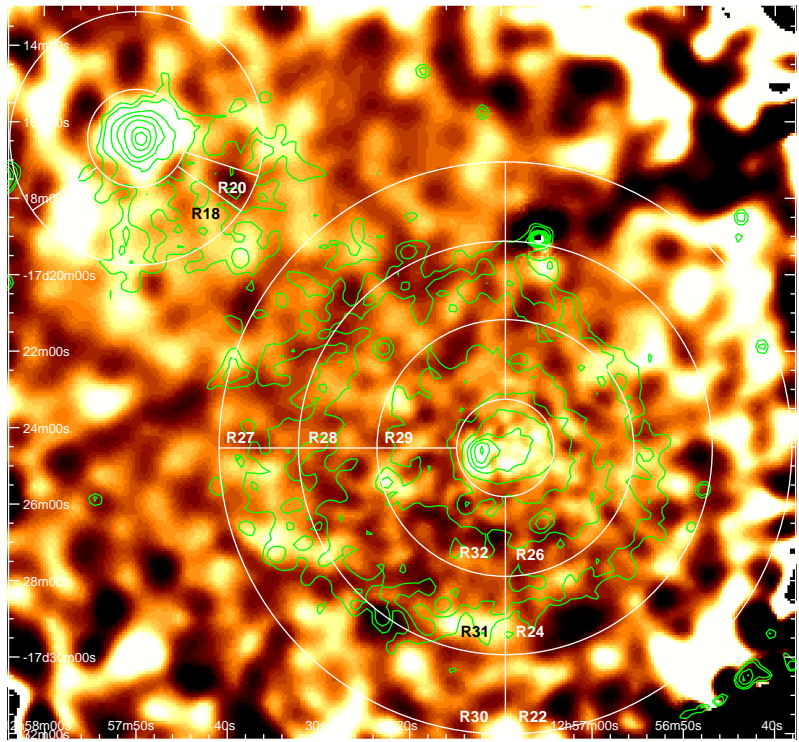}
  \caption{Hardness ratio map of the galaxy cluster A1644.
  Background, exposure, and
  vignetting corrected, adaptively smoothed, combined MOS1-MOS2-pn count rate
  image ratio for the energy bands (0.3--2.0\,keV)/(2.0--10.0\,keV). Soft
  emission appears bright 
  and hard emission dark. Also shown are surface brightness contours for the
  (0.3--2.0\,keV) image and regions selected for spectral analysis.
  \vspace{-0.6cm}} 
\end{figure}

The next step we took is the subdivision of annuli into regular segments.
Figure~2
shows selected regions as well as surface brightness contours overlaid onto a
hardness ratio (HR) map. Preliminary temperature (metal abundance) estimates for
the segments based on standard spectral model fits are shown in the left
(right) hand side of Fig.~3. Now it becomes obvious that the apparently regular
temperature profiles are misleading and only due to averaging over (too) large
regions. The temperatures of the segments to the East of the main clump (R22,
R24, and R26) are all significantly lower than almost all regions to the West
(R27--R32). This may indicate that the gas in the regions between the two clumps
has been heated up by adiabatic compression or even shocks.

Having found irregularities in the temperature structure a further step is to
select regions based on the HR map to directly search for cool/metal rich
trails (bright) or hot spots (dark). The significance of brightness
fluctuations is again
evaluated by direct spectral fits. Note that not all artifacts, e.g.,
inexact exposure correction close to CCD chip boundaries, have been removed in
the HR map in Fig.~2. The spectral analysis, however, is not affected by this.
The region to the South of the sub clump (R18) appears fairly bright and
therefore soft, whereas a small region to the Southeast (R20) appears hard. The
spectral fit results (Fig.~3) reveal that indeed the temperature estimate for
R18 is 
significantly lower than the estimate for the rest of this annulus (R19) and
especially than that for R20. Furthermore the metal abundance estimate for R18
appears enhanced. These findings together with the surface brightness structure
might imply that to the South we see gas that has been
removed from the cooler center of the sub clump possibly by the combined
effect of some energy source related to the central galaxy of the sub clump (a
radio source) and the movement through intracluster gas of the main clump.
Should more detailed modeling confirm the significance of the abundance excess
then other possibilities like cooling of intracluster gas onto the moving sub
clump (as seen in A1795 on much smaller scales, \cite{fse01}) would be more difficult to reconcile with the data.

In summary the temperature structure of A1644 shows that this system
is quite complicated as indicated by the surface brightness structure.
Interestingly no evidence for substructure has been found in optical and
near-infrared observations \cite{tgk01}.
The next step is to attempt an improved mass estimate and compare it to
more simple estimates as generally applied to clusters at higher redshift and in
larger samples. Note that this detailed study required only 12\,ks of good
data.
Chandra and XMM-Newton observations of the complete \gcs\ are quite cheap but
offer a great opportunity for clusters in the era of precision cosmology.
Comparison of cluster X-ray data to the wealth of new multiwavelength data,
e.g., 2MASS, will also help to reduce systematic uncertainties.
\begin{figure}
  \hspace{-0.8cm}
  \includegraphics[height=.25\textheight, angle=0.]{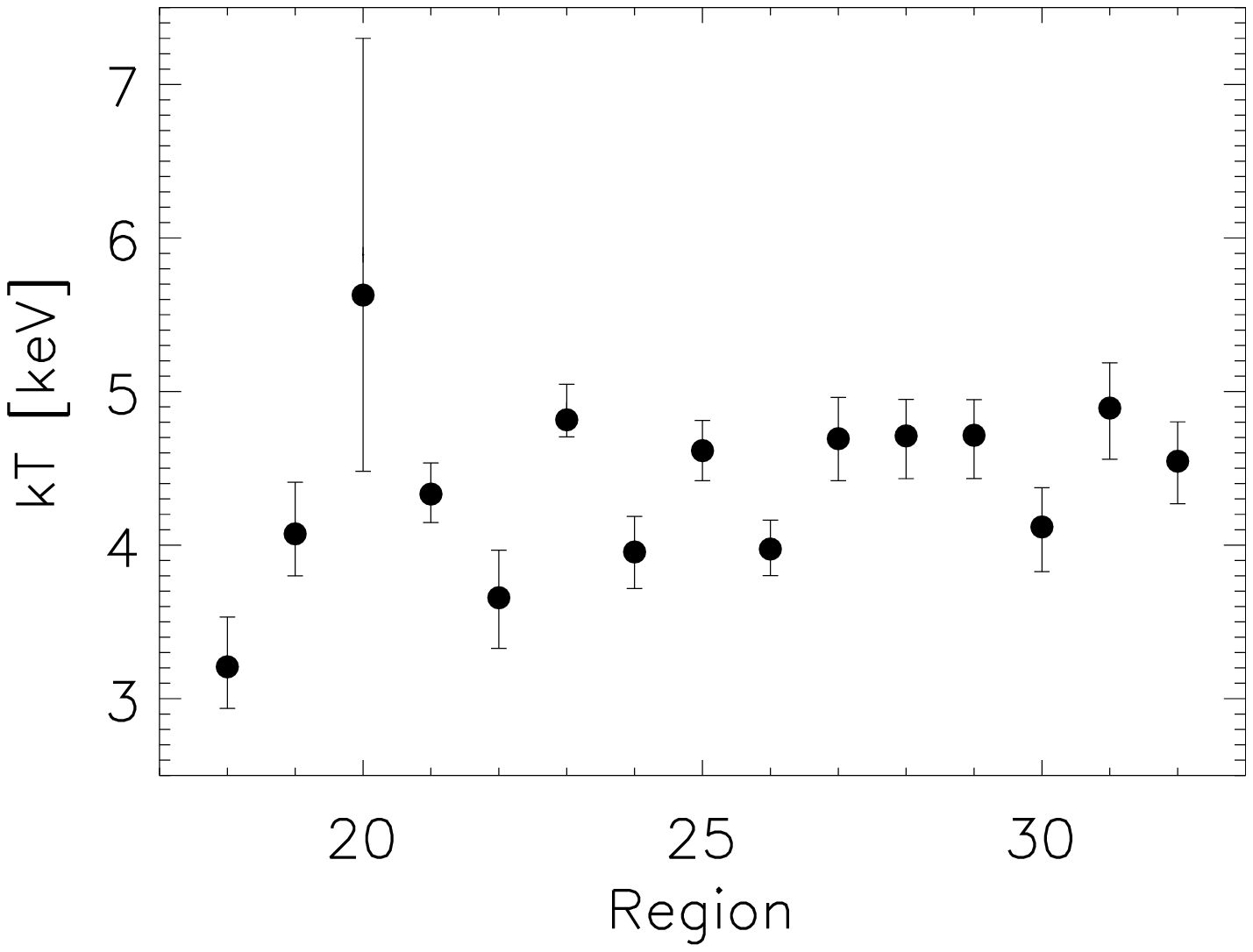}
  \includegraphics[height=.25\textheight, angle=0.]{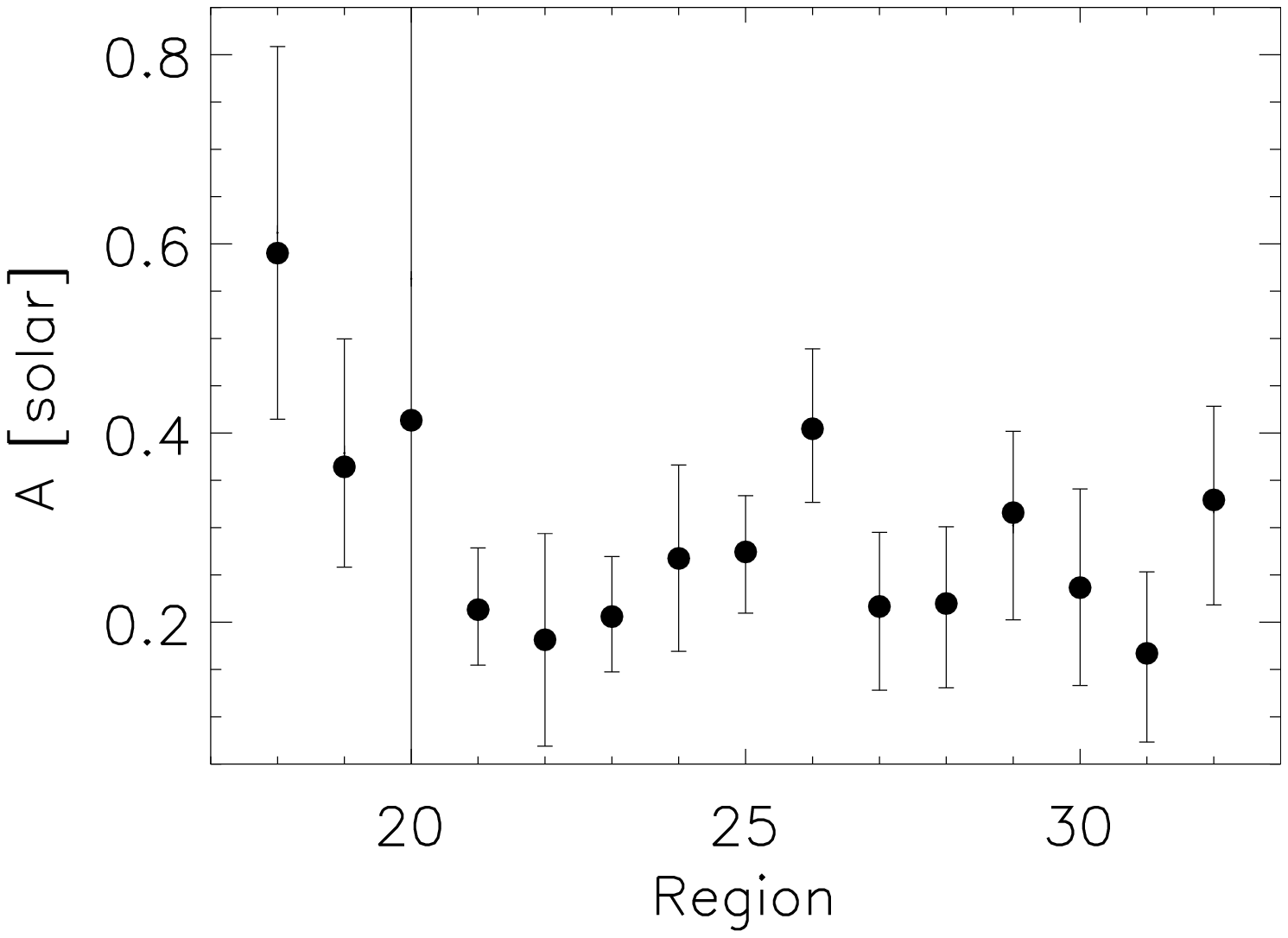}
  \caption{Gas temperature (left) and metal abundance (right) estimates of
  selected regions based on simultaneous spectral fits to background and
  vignetting corrected
  spectra from the MOS1, MOS2, and pn detectors aboard XMM-Newton.
  Note that R21=R27+R30, R23=R28+R31, R25=R29+32, and R19=full annulus--R18;
  see Fig.~2. Statistical
  error bars show the 90\% confidence level for one interesting parameter.
  \vspace{-0.85cm}}
\end{figure}


We acknowledge the benefit from the dedicated work of the NORAS\,II and 2MASS
teams.
This work was supported by NASA XMM-Newton Grant NAG5-10075.
THR acknowledges support by the Celerity Foundation through a 
Post-Doctoral Fellowship.
The XMM-Newton project is an ESA Science Mission with instruments and
contributions directly funded by ESA Member States and the USA (NASA).
This publication makes use of data products from 2MASS, which is a joint project
of the University of Massachusetts and IPAC/Caltech, funded by NASA and NSF.
\vspace{-0.4cm}

\end{document}